\documentclass[aps,prc,reprint,superscriptaddress,twoside,twocolumn,nofootinbib,showpacs]{revtex4-1}

\usepackage{amsmath,amssymb}
\usepackage{braket}
\usepackage{graphicx,bm}
\usepackage{tensor}
\usepackage{siunitx}

\usepackage{multirow}

\usepackage{changes}

\allowdisplaybreaks
\usepackage{color}

\newcommand{\abs}[1]{\left| #1 \right|}

\begin{document}

\title{\boldmath
From homogeneous matter to finite nuclei: Role of the effective mass} 


\author{Hana \surname{Gil}}
\email{gil@knu.ac.kr}
\affiliation{Department of Physics, Kyungpook National University, Daegu 41566, Korea}

\author{Panagiota \surname{Papakonstantinou}}
\email{Corresponding author: ppapakon@ibs.re.kr}
\affiliation{Rare Isotope Science Project, Institute for Basic Science, Daejeon 34047, Korea}

\author{Chang Ho \surname{Hyun}}
\email{hch@daegu.ac.kr}
\affiliation{Department of Physics Education, Daegu University, Gyeongsan, Gyeongbuk 38453, Korea}

\author{Yongseok \surname{Oh}}
\email{yohphy@knu.ac.kr}
\affiliation{Department of Physics, Kyungpook National University, Daegu 41566, Korea}
\affiliation{Asia Pacific Center for Theoretical Physics, Pohang, Gyeongbuk 37673, Korea}

\date{\today}

\begin{abstract}
Recent astronomical observations, nuclear-reaction experiments, and microscopic calculations have placed new constraints on the nuclear equation of state (EoS) and revealed that most nuclear-structure models fail to satisfy those constraints upon extrapolation to infinite matter. A reverse procedure for imposing EoS constraints on nuclear structure has been elusive.
 Here we present for the first time a method to generate a microscopic energy density functional (EDF) for nuclei from a given immutable EoS. 
The method takes advantage of a natural Ansatz for homogeneous nuclear matter, the Kohn-Sham framework, and the Skyrme formalism. 
We apply it to the nuclear EoS of Akmal-Pandharipande-Ravenhall and describe successfully closed-(sub)shell nuclei. 
In the process, we provide predictions for the neutron skin thickness of nuclei based directly on the given EoS. 
Crucially, bulk and static nuclear properties are found practically independent of the assumed effective mass
 value -- a unique result in bridging EDF of finite and homogeneous systems in general.  
\end{abstract}


\maketitle



\section{Introduction}

%
%
The nuclear energy density functional (EDF) represents density functional theory in nuclear physics and provides a unified framework for both finite, self-bound nuclei and the equation of state (EoS) of infinite nuclear matter. 
The founding theorems of density functional theory~\cite{HK64,KS65} were formulated originally for an externally bound, non-degenerate, 
unpolarized electron gas in its ground state, but were followed over the years by numerous generalizations 
to situations including degenerate, polarized, and finite systems, and to excited states, as well as by practical 
justifications and refinements in a hierarchy of approximations~\cite{PS01,Bechstedt}.  
Attempts exist to formalize intrinsic EDFs for self-bound systems~\cite{Engel06,Messud11,Messud12}. 
Formal justifications aside, the framework is widely used in nuclear physics~\cite{BHR03,NMMY16}, 
as exemplified by (though not restricted to) Skyrme models. 
There are various approaches based on the Hartree-Fock approximation, with or without explicit 
correlations beyond mean field or pairing, and extending to linear-response theory.

Current major goals of nuclear EDF research include 
1) establishing connections between EoS parameters and nuclear observables, especially for the purposes of astrophysical modeling and 
2) bridging phenomenological EDF approaches with \textit{ab initio\/} approaches~\cite{BSV2008,Dob2016}, 
which use precise phenomenological two- and three-nucleon potentials~\cite{FP81,APR98,GC09} or  
potentials obtained from effective field theories~\cite{EKLM08, DSS13, DHS15}. 
An important express goal is to provide higher predictive power in the description of exotic nuclei. 
A consistent and realistic description of atomic nuclei and homogeneous nuclear matter 
is necessary in the modelling of neutron stars and supernova simulations. 

After decades of work and hundreds of EDF models, the search for a universal functional is ongoing 
and stumbling blocks remain in reaching the above goals.
Traditional EDF models demonstrate spurious correlations amongst parameters, in particular, 
involving the in-medium effective mass~\cite{BDMNP17,DNMBP17}. 
It is also commonly said that certain observables ``prefer" a low or high effective mass. 
(Respective examples are radii or energies~\cite{BHR03}.)
It turns out that most available models fail to reproduce simultaneously nuclear observables and  reasonably constrained EoS properties~\cite{DLSD12,SGSD12}. 
The situation is certainly unsatisfactory: if an EoS is ``realistic," then by definition it should be able to 
reproduce nuclear properties. 
However, testing a given EoS candidate directly on nuclear properties, without refitting and refining it, has been impossible.

In this manuscript, we address the question: Given a reasonable, based on current knowledge, parameterization for homogeneous matter, 
can we apply it to nuclei with no refitting of the EoS parameters? 
For the first time we 
propose a method to realize such applications and obtain a positive answer to the above question. We find that the key lies in the treatment of the in-medium effective mass. 
Any parameters which do not affect and cannot be constrained from the EoS are fitted to bulk properties of only three nuclei and are found to describe all considered nuclei successfully.
Based solely on the given EoS, predictions for exotic isotopes and for the neutron skin thickness of key nuclei are provided.  

Our method 
 and its first applications to nuclei will be presented below.
In Sec.~\ref{sec:HM} we present for completeness the EoS Ansatz and selected parameterization. 
In Sec.~\ref{sec:FN} we describe how a Skyrme model for nuclear structure calculations can be directly engineered from it and present corresponding results for finite nuclei.  
We summarize in Sec.~\ref{sec:concl}.


\section{Homogeneous matter\label{sec:HM}} As elaborated and justified in Ref.~\cite{PPLH16}, the energy per particle in homogeneous infinite nuclear matter is parameterized in terms of
the Fermi momentum $k_F^{}$ or the cubic root of density,
\begin{equation} \label{eq:KIDS}
\mathcal{E}(\rho,\delta) = \mathcal{T}(\rho,\delta) + \sum_{n=0}^{3} c_n^{} (\delta)\rho^{1+a_n}; \quad a_n=n/3, 
\end{equation} 
where $\rho = \rho_n^{} + \rho_p^{}$ is the baryon density, with $\rho_n^{}$ and $\rho_p^{}$ being the neutron and proton 
densities, respectively, and asymmetry is defined as $\delta \equiv (\rho_n^{} - \rho_p^{})/\rho$. 
The kinetic-energy  part $\mathcal{T}$ is written as 
\begin{equation}
\mathcal{T}(\rho,\delta) 
= \frac{3}{5}  
\left[ 
     \frac{\hbar^2}{2m_p} \Big(\frac{1\! -\! \delta}{2}\Big)^{5/3}  
  \!\!\! 
+ \frac{\hbar^2}{2m_n} \Big(\frac{1\! + \! \delta}{2}\Big)^{5/3} 
\right]  
(3\pi^2\rho)^{2/3} \, ,
\end{equation} 
where $m_p^{}$ ($m_n^{}$) is the proton (neutron) mass. 
The Ansatz (\ref{eq:KIDS}) and related strategy are henceforth dubbed KIDS (Korea: IBS-Daegu-SKKU) after the locale or institute of 
the original developers~\cite{PPLH16,GPHPO16, GOHP17}.
The statistical analyses of Ref.~\cite{PPLH16} showed that three terms suffice for isospin-symmetric nuclear matter (SNM) 
(it is worth noting that the same is concluded in Ref.~\cite{BFJPS17} based on nuclear data) in a converging hierarchy, and that four terms suffice for pure neutron matter (PNM) 
in a broad regime of densities. 
A larger number of parameters is undesirable, as it might lead to overfitting. 

A set of SNM parameters $c_i(0)$ was determined by using established properties 
at saturation: the saturation density $\rho_0^{} = 0.16~\mbox{fm}^{-3}$, the energy per particle at saturation 
$\mathcal{E}_0 = -16~\mbox{MeV}$, and the compression modulus $K_0 = 240~\mbox{MeV}$. 
This information can uniquely fix three unknowns $c_{0,1,2}^{}(0)$. (Because it gives
a marginal contribution, one can set $c_3^{}(0) = 0$~\cite{PPLH16,GPHPO16,GOHP17}.
If necessary, one can use this free parameter to also fix the skewness parameter $Q_0$.)

The four PNM parameters $c_i(1)$ at present have been fitted to the Akmal-Pandharipande-Ravenhall (APR) EoS~\cite{APR98}
 in the density range 0.02-0.96 fm$^{-3}$. 
We note that, although all points available in the APR EoS were used in the fit, a properly chosen cost function ensured that the lower-density regime carried more weight in the fit.
Thus was generated the set of parameters called ``KIDS-ad2" in Ref.~\cite{PPLH16}, 
which has $c_0^{}(0) = -664.52$, $c_1^{}(0) = 763.55$, $c_2^{}(0) = 40.13$, $c_3^{}(0) = 0.0$,  $c_0^{}(1) = -411.13$, $c_1^{}(1) = 1007.78$, $c_2^{}(1) = -1354.64$, $c_3^{}(1) = 956.47$, 
where  all $c_n^{}(\delta)$ values are given in units $\mbox{MeV}\cdot\mbox{fm}^{n+3}$. 
The resulting rounded values of the symmetry energy and its slope, curvature, and skewness at saturation are 
$(J,L,K_{\mathrm{sym}},Q_{\mathrm{sym}})=(33,50,-160,590)$~MeV.  

\begin{figure}
 \centering
   \includegraphics[width=0.99\columnwidth]{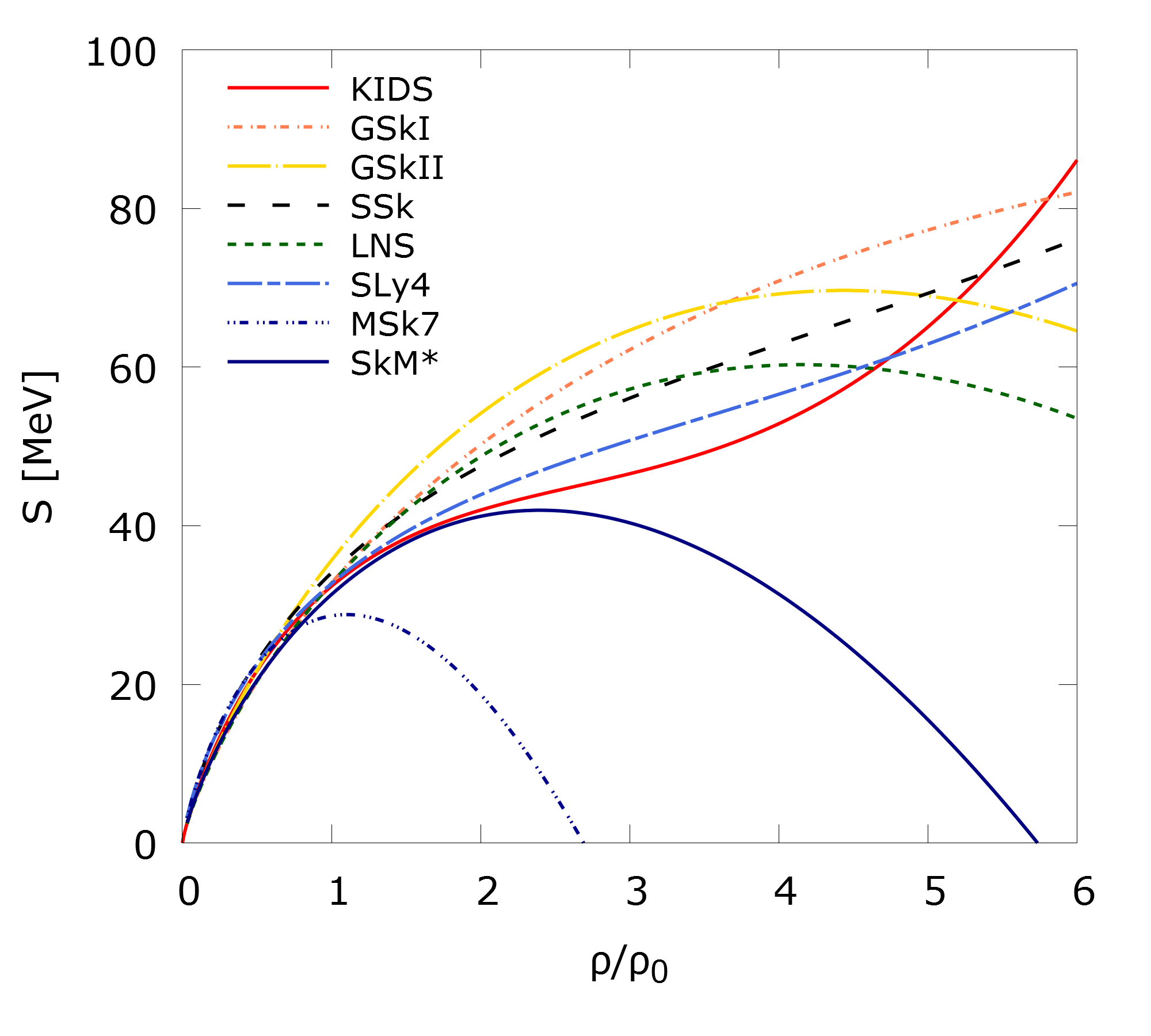}
 \caption{Nuclear symmetry energy. The KIDS-ad2 equation of state employed in this work is 
compared with other available models which have been fitted to nuclear data under various protocols.}
  \label{fig:sym}
\end{figure}

Focusing on high densities, in Ref.~\cite{PPLH16}, the efficiency of the scheme was demonstrated 
in the regime of neutron stars, while the convergence of the expansion was explored Ref.~\cite{GOHP17}.
The nuclear symmetry energy is calculated as shown in Fig.~\ref{fig:sym}. 
The results show that the symmetry energy may not be soft or stiff but may have a nontrivial density dependence.
Interestingly, a similar behavior is seen in Refs.~\cite{LPR10,PKLMR17} originating 
from skyrmion--half-skyrmion phase transition.
Eq.~(\ref{eq:KIDS}) is rich enough to accommodate such behavior. 

\begin{figure}
  \centering
  \includegraphics[width=\columnwidth]{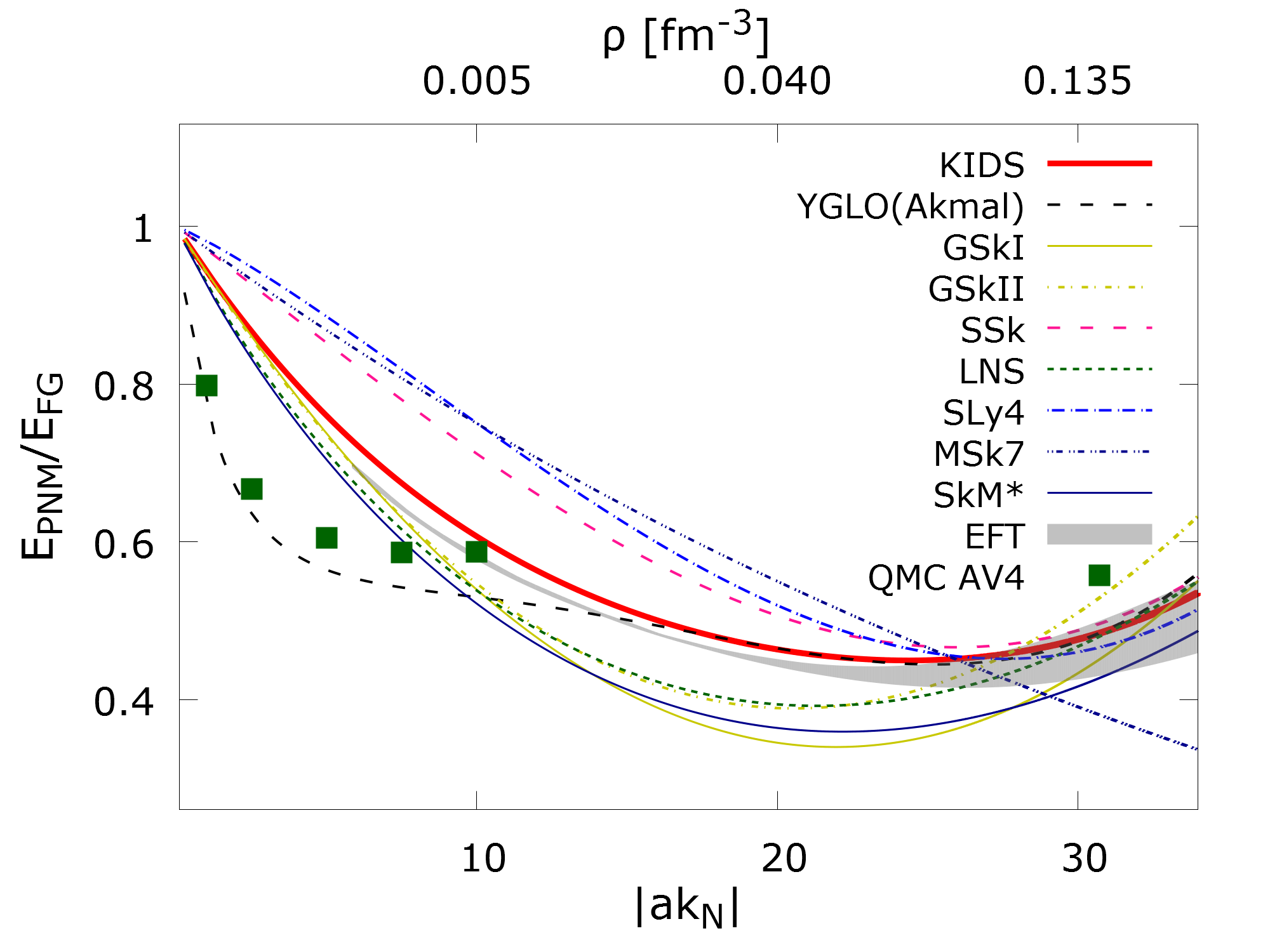}
  \caption{Energy of pure neutron matter $E_{\rm PNM}$ divided by the free-gas value $E_{\text{FG}}$ as a function of $|ak_N|$, 
  where $a(=-18.9$ fm) is the neutron-neutron scattering length in free space, and $k_N^{}$ is the neutron Fermi momentum. 
Corresponding density values are indicated on the upper abscissa. 
The KIDS-ad2 equation of state (full line) is compared with the results of chiral EFT (gray band), a QMC calculation (points), the EFT-inspired model YGLO(Akmal) (black dashed line) and other available models fitted to nuclear data under various protocols.  
}
  \label{fig:Energy}
\end{figure}

Very relevant for nuclei, on the other hand, is the behavior at low densities.
We test the validity of extrapolation below the APR pseudodata region, $\rho < 0.02$~fm$^{-3}$.
The results for the PNM energy are shown in Fig.~\ref{fig:Energy}, where they are compared with EoSs obtained from \textit{ab initio\/} methods 
(QMC AV4~\cite{GC09} and EFT~\cite{DSS13}), from a resummation formula YGLO~\cite{YGL16} 
(fitted to \textit{ab initio\/} pseudodata in $\rho < 0.005$~fm$^{-3}$~\cite{GC09} and to the PNM EoS of Ref.~\cite{APR98}
at $\rho > 0.02$~fm$^{-3}$) and from representative Skyrme models~\cite{DLSD12}.
The KIDS model is found to reproduce the low-density curvature best with respect to EFT, {\em to which it has not been fitted}. 
 
Having shown that the above EoS has a wide range of applicability in density, we apply it to investigate nuclei 
\textit{without altering its given parameter values\/}. 
For the purpose of testing further this ``proof of principle" 
we use additionally a similarly obtained EoS but with a compression modulus of $K_{0}=220$~MeV. 
%
%




\section{Finite nuclei\label{sec:FN}} 

For the application to nuclei we rely on the Kohn-Sham framework.  
In particular, we reverse-engineer a Skyrme-type force~\cite{DLSD12} for Hartree-Fock calculations, which can then be undertaken with a straightforward extension of a standard numerical code~\cite{Reinhard91}. 
Minimally, in order to reproduce the EoS of Eq.~(\ref{eq:KIDS}), we adopt the form of generalized Skyrme force given as 
\begin{equation} 
\begin{gathered} 
v_{ij}^{}  
= (t_0^{} + y_0^{} P_\sigma) \delta(\bm{r}_{ij}^{})  +\frac{1}{2} (t_1^{} + y_1^{} P_\sigma)[\delta(\bm{r}_{ij}) \bm{k}^2 + \mbox{h.c.}]
 \mbox{} \\  
  + (t_2^{} + y_2^{} P_\sigma)\bm{k}^\prime \cdot \delta(\bm{r}_{ij}) \bm{k} 
  + iW_0\,\bm{k}^\prime \times \delta(\bm{r}_{ij})\,
\bm{k}  \cdot (\bm{\sigma}_i^{} +\bm{\sigma}_j^{})
 \\ 
+ \frac{1}{6} \sum_{n=1}^{3}(t_{3n}^{} + y_{3n}^{} P_\sigma) \rho^{a_n} \delta(\bm{r}_{ij})
; \quad a_n=n/3 	
\end{gathered} \label{eq:skyrme}	
\end{equation}
where $\bm{k} = (\bm{\nabla}_i - \bm{\nabla}_j)/(2i)$ and $\bm{k}^\prime = -(\bm{\nabla}_i^\prime - \bm{\nabla}_j^\prime)/(2i)$, 
$\bm{r}_{ij}$ is the relative coordinate, and $P_\sigma$ is the spin-exchange operator.
The strength of the spin-orbit coupling, which is absent in the EoS of Eq.~(\ref{eq:KIDS}), is introduced by the $W_0$ term.
It should be noted that the above ``force" is an auxiliary entity, with no direct relation to a true Hamiltonian. (It is rather used 
as a stepping stone to the equivalent independent-particle ``external" potential $V$ and resulting Kohn-Sham orbitals.)  
Indeed, it is fitted rather than derived.
 More to the point, unless the values of $a_n$ in the density dependence are integer, it cannot consistently be interpreted as a multi-body interaction~\cite{EKR2010}.

The Skyrme force of Eq.~(\ref{eq:skyrme}) resembles other generalized Skyrme models with multiple density-dependent 
couplings $t_{3n}$~\cite{CBBDM04,ADK06,XPC15}.  
However, our strategy for determining not only the precise form~\cite{PPLH16}  but also the strength of the Skyrme parameters 
is completely different. In particular, we keep the given EoS unchanged. 
In addition, we can retain the flexibility to assume arbitrary values for the nucleon effective mass if desired. 

By comparing Eq.~(\ref{eq:KIDS}) and the EDF corresponding to the above Skyrme force~\cite{BHR03,DLSD12},
relations among Skyrme and KIDS parameters can be straightforwardly obtained as
\begin{equation}
\begin{gathered} 
 \label{eq:relation} 
t_0^{} = \frac{8}{3} c_0^{} (0) \, ,  \quad  
y_0^{}=\frac{8}{3} c_0^{}(0) - 4 c_0^{}(1),   \\  
t_{3n}^{} = 16 c_n^{} (0) \, , 
\quad y_{3n}^{} = 16 c_n^{} (0) - 24 c_n^{} (1), \quad (n \neq 2)   \\ 
t_{32}^{} = 16 c_2^{} (0) - \frac{3}{5} \left(\frac{3}{2}\pi^2 \right)^{2/3} \theta_s,  \\ 
y_{32}^{} = 16 c_2^{}(0) - 24c_2^{}(1) + \frac{3}{5}(3\pi^2)^{2/3} \left( 3\theta_{\mu} - \frac{\theta_s}{2^{2/3}}\right),
\end{gathered} 
\end{equation} 
with
\begin{equation} 
\theta_s \equiv 3t_1^{} + 5 t_2^{} + 4y_2^{} \, , \quad \theta_{\mu} \equiv  t_1^{} + 3t_2^{} - y_1^{} + 3y_2^{} \, .
\label{eq:theta} 
\end{equation} 
This reveals that 1) most Skyrme parameters $(t_0,y_0,t_{31},y_{31},t_{33},y_{33})$ are uniquely determined from the KIDS EoS parameters with $n\neq 2$, but 2) the above EDF for nuclei provides two sources for the $c_2(\delta )\rho^{5/3}$ term ($n=2$):
one from the density-dependent term in Eq.~(\ref{eq:skyrme})  (Skyrme parameters $t_{32}^{}, y_{32}^{}$), and the other
from the momentum-dependent terms in Eq.~(\ref{eq:skyrme})  (parameters $t_1^{}, t_2^{}, y_1^{}, y_2^{}$).
The latter parameters determine the isoscalar and isovector effective masses as~\cite{CBHMS97} 
\begin{eqnarray} 
\mu_s^{-1}\equiv ({m^{\ast}_{\mathrm{IS}}/m})^{-1} &=& 1+\frac{m}{8\hbar^2}\, \rho\, \theta_s , 
\\ 
\mu_v^{-1}\equiv ({m^{\ast}_{\mathrm{IV}}/m})^{-1} &=& 1+ \frac{m}{4\hbar^2}\, \rho \, (\theta_s-\theta_{\mu})  , 
\label{eq:effm} 
\end{eqnarray} 
where we have used, for simplicity, the average nucleon mass $m$.  


The unknowns to be determined for nuclei, besides $W_0$, are then the momentum-dependent
proportions in $c_2^{}(0)$ and $c_2^{}(1)$ and, correspondingly, the precise values of $t_{1,2}^{}$ and $y_{1,2}^{}$. 
Our procedure is to 
\begin{enumerate} 
\item 
fit the momentum-dependent terms to the energy and charge radius of $^{40}${Ca} (details specified below),  
with $W_0$ initialized to null,
\item 
determine $W_0$ from the energies and radii of the nuclei $^{48}${Ca} and $^{208}${Pb},
\item 
iterate, i.e., examine $^{40}${Ca} with the new value of $W_0$ and again determine $W_0$ anew and so on. 
\end{enumerate} 
It turns out that iteration is largely unnecessary, because the bulk properties of the spin-saturated nucleus $^{40}${Ca} 
are insensitive to $W_0$.

The momentum-dependent part (steps 1, 3 above) can be determined in different ways. 
A simplistic procedure we explored before~\cite{GOHP17,GPHPO16}  is to set $y_1^{} = y_2^{} = 0$, 
and encode the momentum dependence in a single parameter $k$, corresponding to the portion 
of $c_2(\delta)$ assigned to the momentum dependence part. 
The value of $k$ is then determined from the properties of $^{40}${Ca}. 
For KIDS-ad2 one obtains $k=0.111$, corresponding to ($\mu_s^{} = 0.99$, $\mu_v^{} = 0.82$),
and $W_0=108.35$~$\mbox{MeV}\cdot\mbox{fm}^{5}$. This method provides rather limited flexibility.

A preferable way is to retain the freedom in $y_{1,2}^{}$. 
We now have four parameters to be explored instead of two. 
This freedom allows us to explore different values for the effective masses ($\mu_s,\mu_v$) at saturation density and is central to the present work. 
According to Eqs.~(\ref{eq:relation})$-$(\ref{eq:effm}), the values for the effective masses, together with the already-fixed EoS coefficients 
$c_2(0),c_2(1)$, determine the parameters $t_{32},y_{32}$.
This is different from the traditional Skyrme force model \textit{that has a priori $t_{32}=y_{32}=0$, 
making the whole $c_2^{}$ term to be momentum-dependent}, not necessarily a physical assumption. 
We are now left with only two unknowns in each case, namely two linear combinations of $t_1,t_2,y_1,y_2$. 
In the Skyrme functional those correspond conveniently to the isoscalar and isovector gradient coupling coefficients  $C_0^{\rho\Delta\rho}$ and $C_1^{\rho\Delta\rho}$, 
which are inactive in and unconstrained by homogeneous matter, but the fit to nuclei can determine them.   
In practice, the parameter space can be handily constrained by demanding 
1) that $C_1^{\rho\Delta\rho}$ be lower than $50$~$\mbox{MeV}\cdot\mbox{fm}^5$ 
(a handy and loose enough rule of thumb~\cite{HPDBD13}) 
and 
2) that polarized neutron matter remain stable at high densities~\cite{KW94b}. 
From the acceptable combinations we choose the one that gives the best results for $^{40}${Ca}. 
Finally, for each parameter set [i.e., essentially, for each pair of ($\mu_s,\mu_v$) and best fit to $^{40}${Ca}]
we determine  $W_0$ by fitting to the energies and radii of $^{48}${Ca} and $^{208}${Pb}.

In the following applications we employ primarily the KIDS-ad2 parameterization, which is based on the APR EoS and was already presented above,
and we explore the momentum dependence. 
Skyrme parmeters for different values of the effective masses are derived, as already described, for: 
\begin{itemize}    
    \item 
         The KIDS-ad2 EoS for SNM and PNM 
         and for $\mu_s^{}=0.7,0.8,0.9,1.0$ (with $\mu_v^{}=0.82$) or 
         for $\mu_v^{}=1.0$ (with $\mu_s^{}=0.9$).  
   \item 
         For verification purposes, the same EoS parameters as KIDS-ad2 except that the SNM compression modulus is $K_{0}=220$~MeV; 
 thus $c_0^{}(0) = -727.02$, $c_1^{}(0) = 993.80$, $c_2^{}(0) = -171.93$, $c_3^{}(0) = 0.0$, where $c_n$ values are in units of MeV$\cdot$fm$^{n+3}$.  
\end{itemize} 
For comparison, we apply also the Skyrme parameter set with $y_1=y_2=0$ and $k=0.111$~\cite{GOHP17,GPHPO16}, henceforth labeled KIDS0:   
\begin{itemize} 
    \item 
         KIDS0: The same EoS as KIDS-ad2, but with  
         $ y_1^{}=y_2^{}=0$, $\mu_s^{}=0.99$, $\mu_v^{}=0.82$ , $W_0=108.35$~MeV.
\end{itemize} 
Resulting Skyrme-type parameters for representative cases 
are collected in Table~\ref{table1}.

\begin{table*}
\renewcommand{\arraystretch}{1.2} 
\centering
\begin{tabular}{c|rrrrrrccc}
\hline \hline
	Model, or: & & & & & & & & & 
$^{60}${Ca}:
\\
 $K_0$ 
	& $t_0^{}$ 			& $t_1^{}$		& $t_2^{}$		& $t_{31}^{}$	& $t_{32}^{}$	& $t_{33}^{}$
	& \multirow{2}{*}{$W_0$} 	& $D_E^{} [\%]$	&  $\frac{E}{A}$~[MeV]\\	
 $(\mu_s,\mu_v$)	& $y_0^{}$			& $y_1^{}$ 		& $y_2^{}$ 		& $y_{31}^{}$	& $y_{32}^{}$	& $y_{33}^{}$
	& $$			& $D_R^{}[\%]$	& $R_c$ [fm]  \\ \hline
\multirow{2}{*}{KIDS0}
	& $-1772.04$	& $275.72$	& $-161.50$	& $12216.73$	& $571.07$	& $0$		
	& \multirow{2}{*}{$108.35$}	& $0.32$		& $7.6561$ \\
	& $-127.52$		& $0$		& $0$		& $-11969.99$	& $29485.49$	& $-22955$	
	&  $$		& $0.56$		& $3.6465$	
	\\ \hline
{$240$~MeV}
	& $-1772.04$	& $270.52$	& $-355.95$	& $12216.73$	& $642.12$	& $0$		
	& \multirow{2}{*}{$97.61$} 	& $0.38$		& $7.6993$ \\
{$(1.0, 0.82)$}	& $-127.52$		& $156.90$	& $242.04$	& $-11969.99$	& $29224.07$	& $-22955$	
	& $$			& $0.56$		& $3.6416$	
	\\ \hline
{$240$~MeV}  
	& $-1772.04$	& $448.99$	& $-279.45$	& $12216.73$	& $-2572.65$	& $0$		
	& \multirow{2}{*}{$135.24$}	& $0.26$		& $7.6464$ \\
{$(0.7,0.82)$}  	& $-127.52$		& $-345.72$	& $234.74$	& $-11969.99$	& $41318.69$	& $-22955$	
	& $$		& $0.52$		& $3.6420$	
	\\ \hline
{$240$~MeV}
	& $-1772.04$	& $315.97$	& $-527.58$	& $12216.73$	& $-191.34$	& $0$		
	& \multirow{2}{*}{$107.58$}	& $0.38$		& $7.6933$ \\
{$(0.9,1.00)$}	& $-127.52$		& $-56.87$	& $480.10$	& $-11969.99$	& $36289.12$	& $-22955$	
	& $$			& $0.57$		& $3.6370$	
	\\ \hline
{$220$~MeV}
	& $-1938.71$	& $281.04$	& $-479.05$	& $15900.76$	& $-2750.91$	& $0$		
	& \multirow{2}{*}{$88.96$}		& $0.52$		& $7.7701$ \\
{$(1.0,0.82)$}	& $-294.19$		& $236.07$	& $388.03$	& $-8285.96$	& $25831.04$	& $-22955$	
	& $$		& $0.94$		& $3.6524$	
	\\ \hline
{$220$~MeV}
	& $-1938.71$	& $466.23$	& $-439.68$	& $15900.76$	& $-5965.68$	& $0$		
	& \multirow{2}{*}{$133.36$}	& $0.44$		& $7.6807$ \\
{$(0.7,0.82)$}	& $-294.19$		& $-247.10$	& $422.10$	& $-8285.96$	& $37925.67$	& $-22955$	
	&  $$		& $0.82$	& $3.6663$	
	\\ \hline
\multirow{2}{*}{GSkI~\cite{ADK06}}
	& $-1855.45$		& $397.23$	& $264.63$	& $13858.00$	& $-2694.06$	& $-319.87$	
	& \multirow{2}{*}{$169.57$}	& $0.16$ 		& $7.6294$ \\
	& $-219.02$		& $-698.59$	& $-478.13$	& $1747.29$	& $3200.69$	& $146.94$	
	&		& $0.50$		& $3.6640$
	\\ 	\hline
\multirow{2}{*}{SLy4~\cite{CBHMS98}}
	& $-2488.91$	& $486.82$	& $-546.39$	& $13777.00$	& $$			& $$			
	& \multirow{2}{*}{$122.69$}		& $0.33$ 		& $7.7030$ \\		
	& $-2075.75$	&$-167.37$	& $546.39$	& $18654.06$	& $$			& $$			
	& $$			& $0.91$ 		& $3.6734$
	\\   \hline
\hline
\end{tabular}
\caption{The Skyrme-type parameters, deviations of the calculated energies and charge radii  
defined by Eq.~(\ref{Eq:errornuclei}), and the predictions for $^{60}${Ca}, for the models indicated on the leftmost column.
The corresponding powers of the density-dependent couplings, $(a_1,a_2,a_3)$, are as in Eq.~(\ref{eq:KIDS}), except for SLy4 where there is only one density-dependent term with $a_1=1/6$.  
The errors $D_E$ and $D_R$ in the cases of GSkI and SLy4 are calculated based only on the values reported 
in the respective original publications~\cite{ADK06,CBHMS98}. 
\label{table1}} 
\end{table*}

\begin{figure}
\centering
\includegraphics[width=0.95\columnwidth]{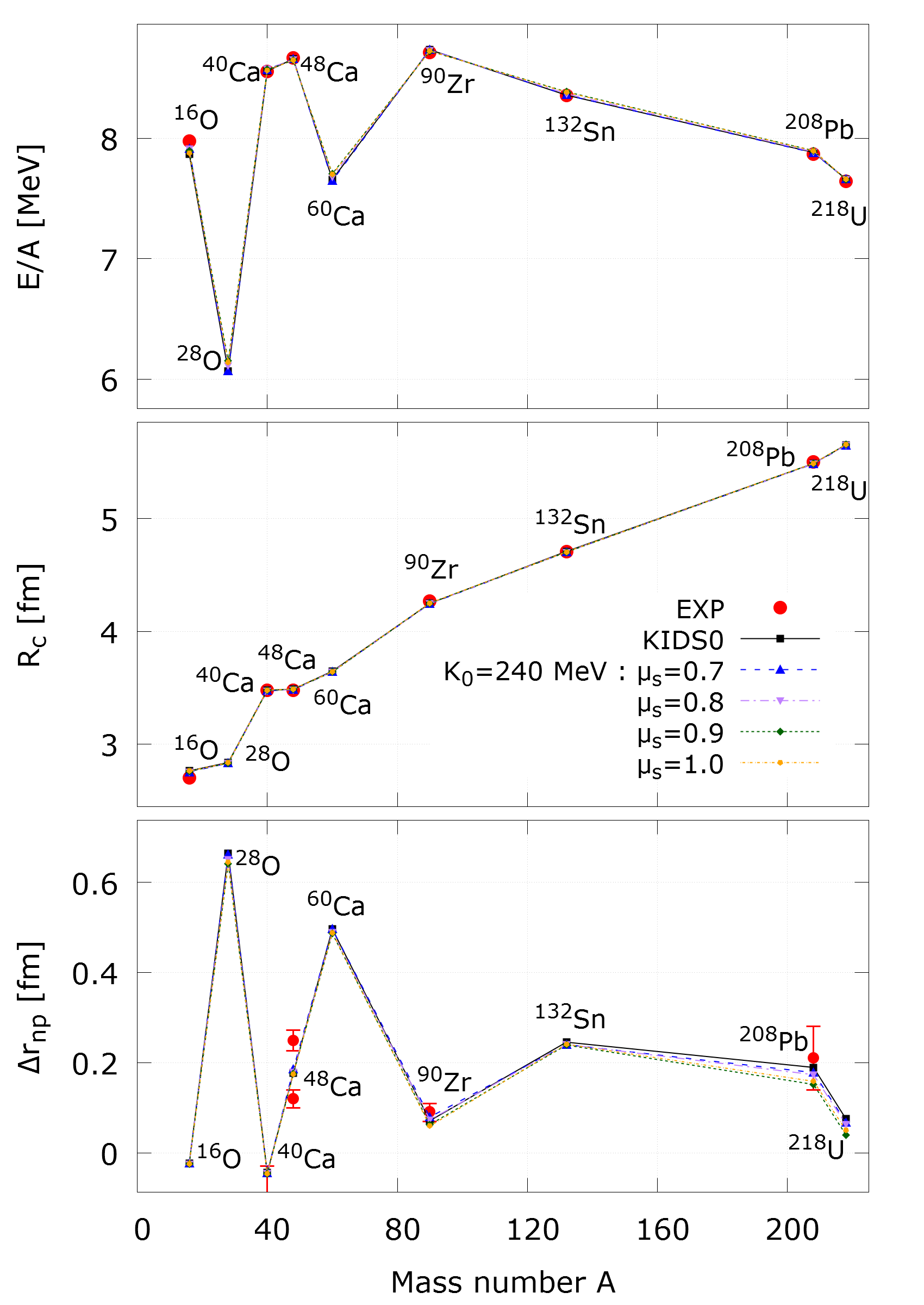}
\caption{Results for binding energy per nucleon $E/A$, charge radius $R_{c}$, and neutron skin thickness $\Delta r_{np}$.
 All results for$^{16}${O}, $^{28}${O}, $^{60}${Ca}, $^{90}${Zr}, $^{132}${Sn}, 
and $^{218}${U} and all neutron-skin results are predictions. 
Data of energy per particle and charge radius are taken from National Nuclear Data Center and Ref.~\cite{AM13}, while neutron skin data from 
Refs.~\cite{JTLK04,MACD17,PREX12}.}
\label{Fig:Nuclei}
\end{figure}

Results from Hartree-Fock calculations for the input nuclei $^{40,48}${Ca}, $^{208}${Pb} and for other (semi-)magic nuclei are 
shown in Fig.~\ref{Fig:Nuclei} along with the available data.   
Here, all results for $^{16}${O}, $^{28}${O}, $^{60}${Ca}, $^{90}${Zr}, $^{132}${Sn}, 
and $^{218}${U} are predictions. 
We compute a mean absolute deviation of the calculated observable $O$ ($O=E/A$ or $R_c$) with respect to data defined as
\begin{equation}
D_O = \frac{1}{N_{\rm nucl}}\sum_{i=1}^{N_{\rm nucl}}  \left|  
\frac{O_{i}^{\mathrm{expt}}-O_{i}^{\mathrm{cal}}}{O_{i}^{\mathrm{expt}}}
\right| \, , 
\label{Eq:errornuclei}
\end{equation}
where the sum runs over nuclei considered here for which data exist.
Results are shown in Table~\ref{table1}.

We observe that in the scale of the graphs the results for the bulk properties are practically indistinguishable 
and, apart from $^{16}${O}, in excellent agreement with available data and on a par with other models 
as the values of $D_E$ and $D_R$ indicate. 
From these results, we draw two conclusions: 
(1) \textit{A good-quality Skyrme model can easily be reverse-engineered from a good-quality EoS without refitting the latter\/},
while (2) \textit{bulk and static quantities are practically independent from the effective mass\/}.
The second conclusion is a most unusual, though intuitively unsurprising, result, which   
would not have been revealed had we ascribed the $\rho^{5/3}$ term fully to the kinetic energy from the outset.
Note that, in that case, for KIDS-ad2 we would have obtained $\mu_s=0.92$ and $\mu_v=0.33$.

The insensitivity to the effecitve mass assumptions is also examined in Table~\ref{table1} for the exotic nucleus $^{60}$Ca which was recently discovered~\cite{Tar2018}. 
Some dependence is perceived in the much finer cases of the neutron skin thickness, in particular, 
of $^{208}${Pb}, $^{218}${U}, and $^{90}${Zr}, possibly attributable to structural details that should be 
examined further in subsequent studies.  
We also observed a tendency that the $K_0=240$~MeV parameterizations perform better than those with $K_0=220$~MeV. 
Systematic studies will be reported elsewhere.

\begin{figure}
  \centering
  \includegraphics[width=\columnwidth]{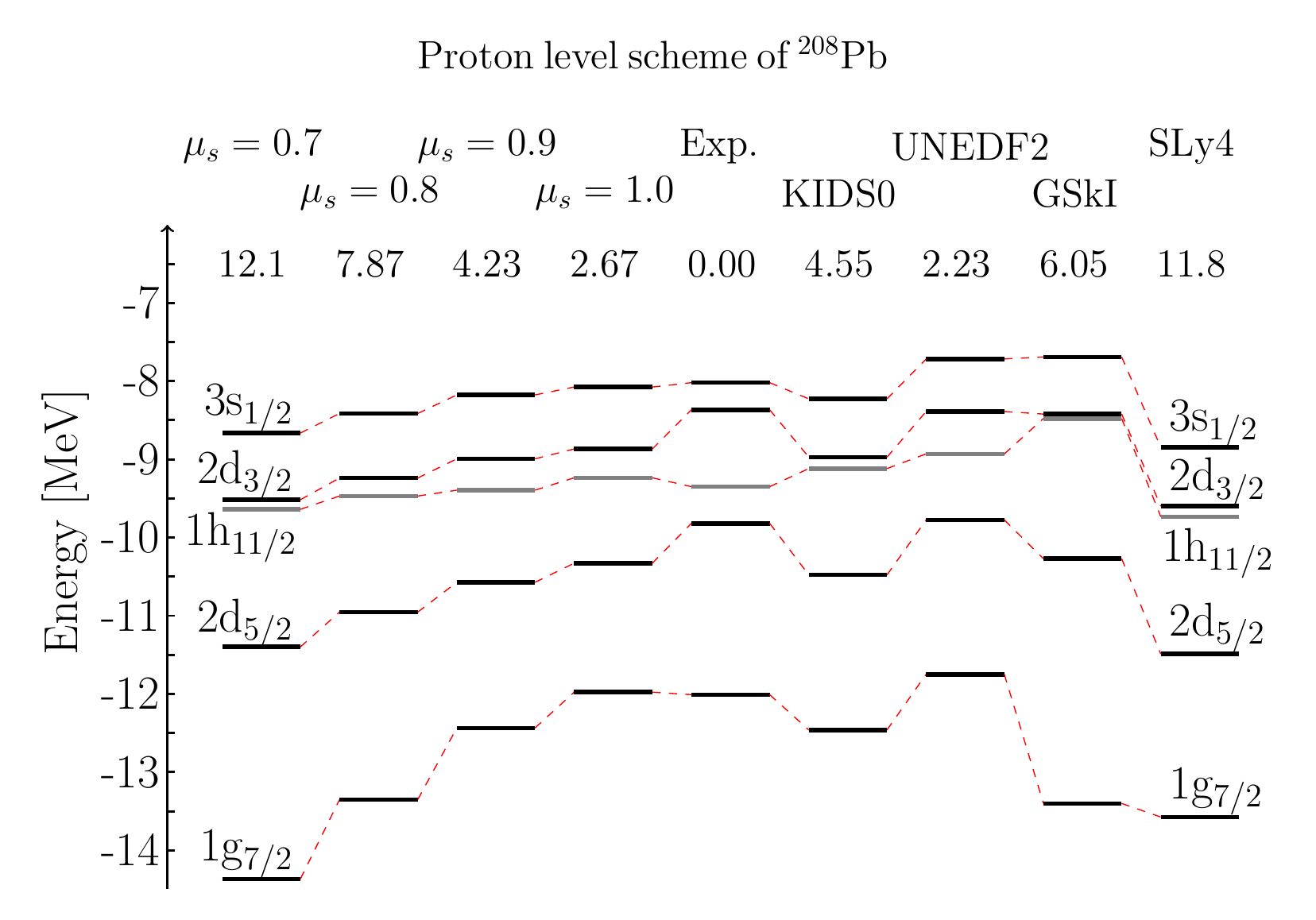}
  \caption{Energies of occupied proton levels of $^{208}${Pb} from various  models 
  compared with empirical removal energies~\cite{SWV07}. 
  The KIDS results for $K_{0}=240$~MeV, $\mu_v=0.82$ and varying $\mu_s^{}$ are on the left of the experimental levels.
 The number under each model name is the deviation $D$ of Eq.~(\ref{eq:md}), in percentage points, for the levels shown underneath. 
}
  \label{Fig:Levels}
\end{figure}

The effective mass values can of course affect dynamical properties such as removal and capture energies 
(single-particle spectrum) and nuclear collective motion. 
Fig.~\ref{Fig:Levels} compares presently obtained single particle energies with data and other model calculations, 
UNEDF2~\cite{KMNO13}, GSkI~\cite{ADK06}, and SLy4~\cite{CBHMS98}.
In UNEDF2 and GSkI the single particle energy levels of $^{208}${Pb} are used in the fitting, so
they are expected to give better results. 
For a quantitative comparison we consider a mean absolute deviation,
\begin{eqnarray}
D \equiv 
\frac{1}{N} \sum_{i=1}^{N} \abs{\frac{E_i^{\text{expt}} - E_i^{\text{cal}}}{E_i^{\text{expt}}}} 
\label{eq:md}
\end{eqnarray}
with the sum executed on all states shown in Fig.~\ref{Fig:Levels}.  
This value is given in units of percentage points under the name of each model in Fig.~\ref{Fig:Levels}. 
The accuracy of KIDS models with high effective mass is similar to those of GSkI and UNEDF2 models.
Strictly speaking, the energy of only the highest occupied state of a many-body system can be considered as 
an observable~\cite{Bechstedt,DV}. 
The comparison is nonetheless interesting, in confirming that higher values of $\mu_s^{}$ may be needed 
to reproduce the single-particle spectrum of $^{208}${Pb}.

\begin{figure}
  \centering
 \includegraphics[width=0.75\columnwidth]{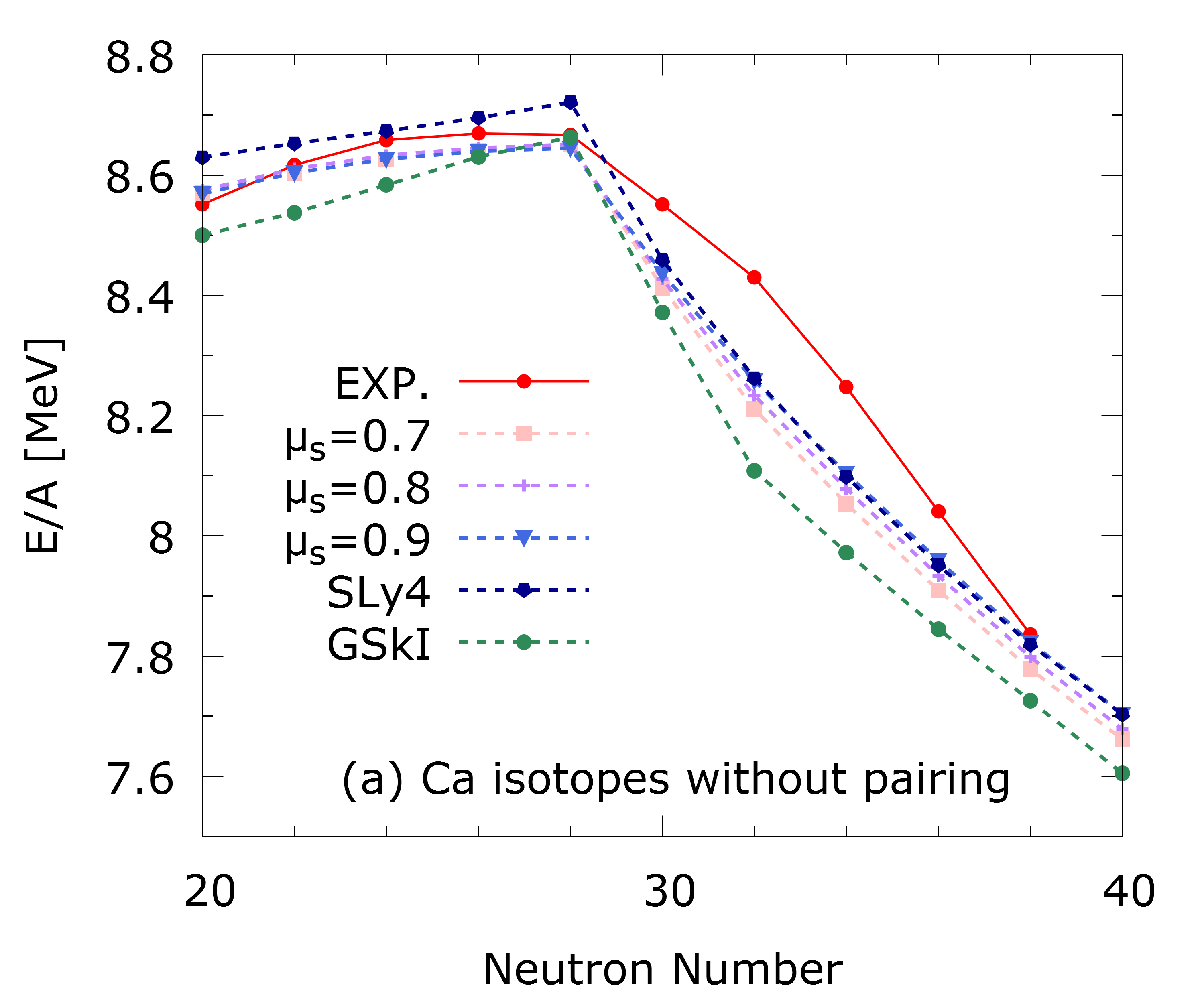} \includegraphics[width=0.75\columnwidth]{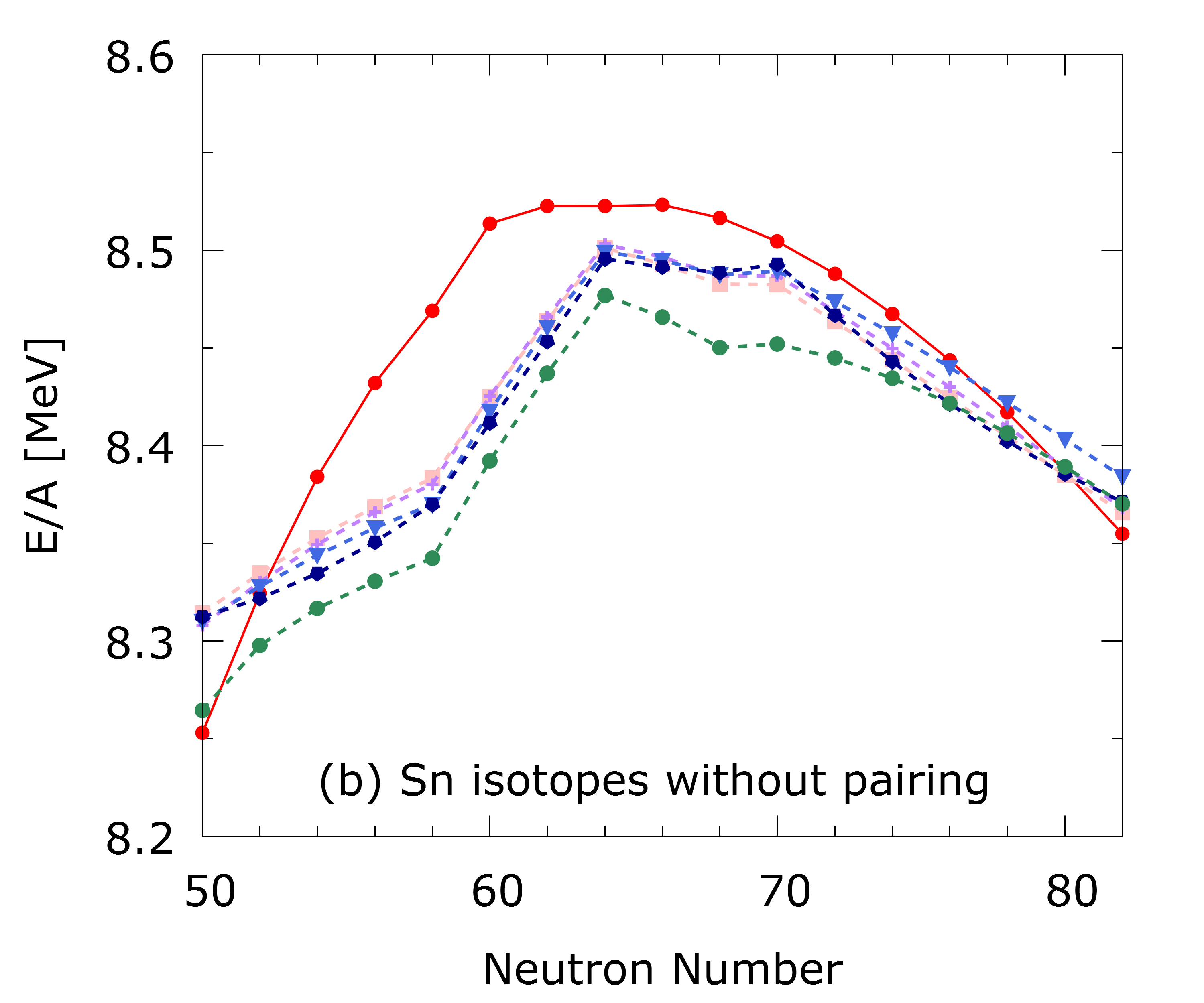}
  \caption{Energy per particle of (a) Ca and (b) Sn isotopes without considering pairing correlations.
  The KIDS results for $K_{0}=240$~MeV, $\mu_v=0.82$ and varying $\mu_s^{}$ are shown along with 
those from representative Skyrme functionals (SLy4, GSkI) and experimental data. 
}
  \label{Fig:Open}
\end{figure}

Since the effective mass value influences the level density, it is relevant when one considers an effective pairing force (unless perhaps consistency issues are properly addressed~\cite{GSMS1999}). 
The latter is active in open-shell nuclei. 
The issue goes beyond the scope of the present work, but two remarks are in order at present. 
First, if closed-shell nuclei are well described by KIDS, then an effective pairing treatment with KIDS functionals can be at least as accurate as with other functionals in use. 
Second, the flexibility KIDS retains in the values of the effective mass will lead to more flexibility when examining pairing. 
As an illustration of the above, in Fig.~\ref{Fig:Open} we show the energy per particle in Ca and Sn isotopes, without pairing, calculated with the KIDS-ad2 parameterization for different values of the effective mass and with two representative Skyrme functionals, along with data. 
The trends along isotopic chains are the same for all functionals.  In addition, a residual effect of the effective mass is seen in some open-shell isotopes. A study of pairing with KIDS in a Hartree-Fock-Bogolyubov (HFB) framework will be reported elsewhere.

\section{Summary\label{sec:concl}} 

We presented and validated a unique method for extracting a generalized Skyrme-type EDF for nuclei from a given, immutable EoS.  
The scheme utilizes a natural and versatile Ansatz for the EoS of SNM and PNM which is ``agnostic" with respect to effective masses.  
We have shown that 
1) a predictive Skyrme model can easily be reverse-engineered from a realistic EoS (one that respects generally accepted constraints around the saturation point) without refitting the latter (a need for refitting signifying an unrealistic EoS by definition), 
and that 2) bulk and static quantities are practically independent of effective masses.
To our knowledge these are unique results unifying the description of finite and homogeneous systems.  
Future applications abound: 
Our method will allow us to vary independently and at will all relevant EoS parameters (e.g., compression modulus or symmetry energy parameters) around a baseline set of values (here, those of KIDS-ad2) and the effective mass and to examine their effects on predictions for nuclear observables, 
with special focus on exotic nuclei to be explored in new rare-isotope facilities~\cite{JPI2018}.
An exploration of symmetry-energy parameters is underway~\cite{Ahn18,AhP18}. Explorations of giant resonances and related issues regarding the role of the EoS parameters and effective mass~\cite{BAS2018}, are also in progress~\cite{PaG2018}. 
It remains feasible and a future goal to constrain the momentum and spin dependence based on microscopic calculations of the 
effective mass or polarized matter and the spin-orbit coupling from, e.g., relativistic approaches.


\section*{Acknowledgments} 
We would like to thank M.~Pearson and A.~Richter for useful comments on the manuscript and B.~K. Agrawal for providing us with the final values of GSkI and GSkII parameters. 
This work was supported by the National Research Foundation of Korea under Grant 
Nos. NRF-2015R1D1A1A01059603 and NRF-2017R1D1A1B03029020. 
The work of P.P. was supported  by the Rare Isotope Science Project of the Institute for Basic Science funded 
by Ministry of Science, ICT and Future Planning and the National Research Foundation (NRF) of Korea (2013M7A1A1075764).

\end{document}